\documentclass[aps,pre,reprint]{revtex4-1}
\usepackage{graphicx}
\usepackage{dcolumn}
\usepackage{bm}
\usepackage{hyperref}

\draft 

\begin{document}




\title{Semi-quantitative predictions for the Polish parliamentary elections (October 25, 2015) based on the Emotion/Information/Opinion model} 
\author{Pawel Sobkowicz}
\email[]{pawelsobko@gmail.com}
\affiliation{KEN 94/140, 02-777 Warsaw, Poland}


\date{\today}

\begin{abstract}
Based on an agent based model of opinion changes,  described in detail in a recent paper (\cite{sobkowicz2015breakdownV2}, \url{arXiv:1507.00126}), we attempt to predict, three months in advance, the range of possible results of the Polish parliamentary elections, scheduled for October 25, 2015. The model reproduces semi-quantitatively the poll results for the three parties which dominated the recent presidential elections and allows estimation of some variations of the electoral propaganda campaigns by the parties. \end{abstract}

\pacs{89.65.-s, 89.75.Fb, 02.70.-c}


\maketitle 

\section{Introduction}
\label{sec:introduction}

A recently published agent based model of the breakdown of a stable two party political stalemate \cite{sobkowicz2015breakdownV2} allows to consider various scenarios dependent on the propaganda strategies of the political forces. The model itself, considering the interplay of emotions and information in shaping of the individual opinions has been presented  in detail in a series of papers \cite{sobkowicz12-7,sobkowicz13-2,sobkowicz13-3}.

In this short communication we attempt to refine the model parameters (in particular the ones describing the party communication strategies), to achieve a semi-quantitative agreement with the 
data describing the support of political parties in Poland, between January and August 2015, 
as well as to predict the range of outcomes for the forthcoming Polish parliamentary elections.  

The paper is organized as follows: Section II provides a short introduction to the political situation, to provide the context for the model; 
Section III describes the simulation procedure, especially the description of the model for the party propaganda,  which provides the basis for the parameter choices. Section IV  contains the results of the model and the discussion of the dependence on the parameters. We refer the Reader to the references mentioned above for the details of the model, in particular the aspects of agent-agent interactions and the interplay of emotions and information leading to the opinion changes in individual agents.

\section{Political background}
\begin{figure*}
\includegraphics[scale=1.0]{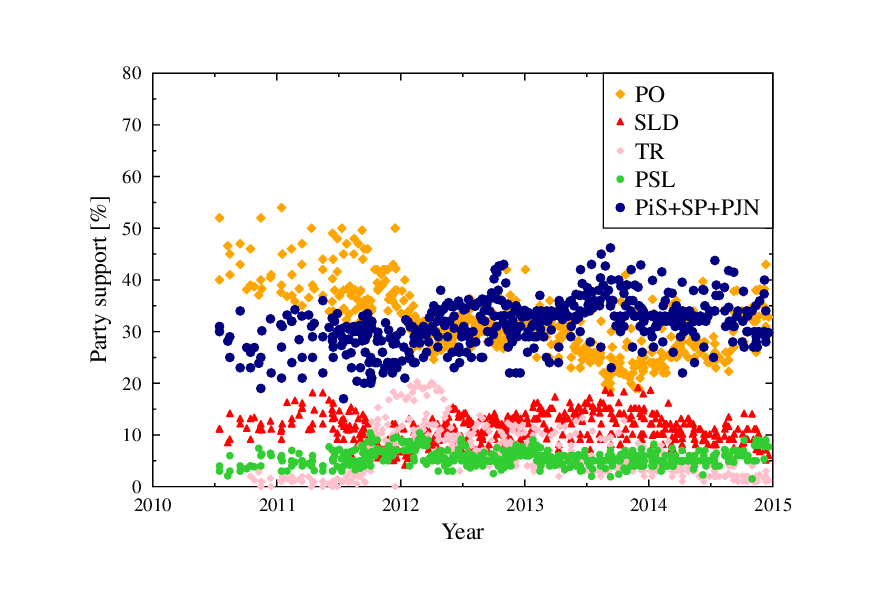}
\caption{Evolution of the support for major political parties in Poland 2010--2014. PO and PiS are the main contenders, while SLD (socialdemocrats) and PSL (peasants party) are examples of smaller parties, existing on the scene during the whole 25 year post-communist era, with small but well entrenched core electoral base. The data on PiS groups them together with two other parties, PJN and SP, which have split-off from PiS during the period, but which have since returned to form a single political entity in late 2014. As the three parties address the same electorate with very similar propositions, we treat them together in the polls analysis.
Since 2007, PSL is the coalition partner of PO. 
TR (Ruch Palikota/Twoj Ruch) is a party formed in mid-2011 by a PO dissident, which has enjoyed a brief period of success between 2011 and 2012.  Data from Mr Maciej Witkowiak, \url{http://niepewnesondaze.blogspot.com/}. \label{fig:long} }
\end{figure*}

\begin{figure*}
\includegraphics[scale=0.9]{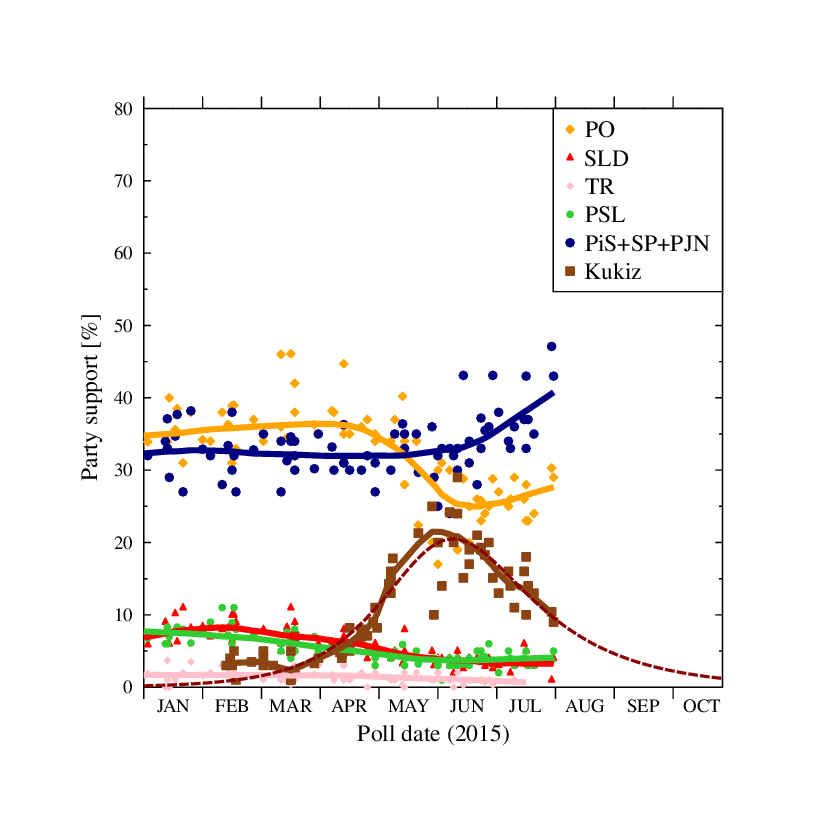}
\caption{Evolution of the support for major political parties in Poland in 2015. The data for Mr Kukiz party (which started to get recognized in the polls since June 2015) is spliced with his personal support in the presidential elections.   Due to the nature of the movement, based largely on the personality of the leader, we think such data use is permissible. Solid lines are lowess smoothed data, while the dashed line for the Kukiz support represents the best fit of a formula combining logistic growth and exponential decay \cite{sobkowicz2015breakdownV2}. Data from Mr Maciej Witkowiak, \url{http://niepewnesondaze.blogspot.com/}.  \label{fig:parties2015} }
\end{figure*}

The political landscape of Poland for the past ten years has been dominated by two parties  Platforma Obywatelska (PO, Civic Platform) and Prawo i Sprawiedliwo\'s\'c (PiS, Law and Justice).
Since 2005 until 2014, the two parties have enjoyed a virtual dominance on the Polish political scene, as documented not only by the results of the parliamentary elections but also the presidential ones, elections to the European Parliament and local elections. 
It is also confirmed by the data from popularity polls, which show practically stable, comparable in size high level support (25\% to 45\%) for the two parties. \footnote{The stability is even more visible if one adds to the results of PiS the figures for two smaller parties: PJN and SP, which have split off from PiS in 2011 and 2012, due to personal differences, keeping a very similar political program and addressing the same electoral base. The two smaller parties merged back with PiS in 2014.} 
During the past five years, the leadership changes a few times, but as can be seen from Figures~\ref{fig:long} and \ref{fig:parties2015}, the differences between the support for each party, as measured by individual polls, is comparable to the difference between the two parties.
There are several other political parties  active on the political scene, but until 2015 their support was much smaller than for the PO and PiS (typically the largest of the minor parties captured, at best, less than half of the support of the two dominant ones). Between 2005 and 2007 the government was formed by a PiS led coalition, but the party was forced to call an early elections in 2007, which it lost to PO.  Since then, a coalition of PO and PSL (peasants' party) has been in power, winning again in 2011, while PiS has remained the strongest opposition party.

A significant change occurred during the first half of 2015, driven by the presidential election campaign (the two tours of the elections were held on May 10 and May 24).  Initially, the contest was  between the incumbent president, Mr Bronislaw Komorowski  (supported by PO) and the candidate of PiS, Mr Andrzej Duda. The initial polls indicated roughly 60\% support for Mr Komorowski, allowing him to win in the first voting round. The situation changed when an independent candidate, Mr Pawel Kukiz, running a grassroots campaign without significant funding or organized support, managed to gather, in the space of a few months, more than 20\% of votes in the first round of the election. This support, driven largely by the dissatisfaction with the current political state, occurred mostly at the expense of the PO candidate. The equilibrium between PO and PiS was broken, and the second round of voting led to the election of Mr Duda. 
Moreover, the increase of the PiS popular support extended beyond the presidential elections, as can be seen from Figure~\ref{fig:parties2015}. At the same time, the support for Mr Kukiz began to drop very fast a few weeks after the elections, despite his declarations of forming a permanent political movement.

A plausible explanation of the apparent ease with which Mr Kukiz gained popularity is based on the asymmetry of the prior communication strategies of the two main parties. Relying on strongly polarized media, which allowed to specify the target of the propaganda messages (with most of the communication addressed to the current supporters), the parties maintained the status quo. 
The crucial difference between the approaches of the two parties was that in the case of PO, most of the messages focused on positive imagery, without invoking strong emotions and allowing room for a rational debate and doubts, the communication strategy of PiS relied on buildup of strong, negative emotions establishing, within their electorate,  the party view as an unquestionable one.

In this situation, the appearance of Mr Kukiz as an alternative was accepted much more easily by the PO electorate. The lack of the emotional commitment allowed to exploit any local dissatisfaction -- in fact many of the voters who supported Mr Kukiz admitted that they voted for him to show PO that they are dissatisfied with some aspects of the 8 years of PO led government. On the other hand, the emotional mobilization of the PiS electorate has allowed them to consolidate their wins and gain momentum ahead of the parliamentary elections in October.

\section{Modeling party propaganda}

The detailed explanation of the model describing how the ``invasion" by Mr Kukiz changed the political landscape was presented in \cite{sobkowicz2015breakdownV2}. In the current work we attempt to extend the model to describe the two months after the presidential elections (June--July) and provide a prediction for the possible outcome of the October parliamentary elections.

The key elements of the model is the description of the factors driving the individual opinion changes.
These may be divided into two parts. The first is related to direct interactions between people, described using the Emotion/Information/Opinion (E/I/O) agent based model, introduced in \cite{sobkowicz12-7,sobkowicz13-2,sobkowicz13-3}. Using an agent based model in a simple 2D geometry it is possible to derive the stable two party outcome, corresponding to the PO/PiS split in the society. 
In the absence of external propaganda, such split may be a very stable one.

The situation may change due to media communications and the propaganda of the parties. 
These differ from the agent-to-agent contacts in the range of operation. While the interpersonal contacts are assumed to be short range (among direct contacts), media may reach large groups simultaneously and act within and outside the enclaves of local consensus. They also reflect, in an organized way, the communication strategies of the parties.

The proper treatment of the propaganda messaging  within the model requires the combination of conceptual simplicity (to allow at least qualitative understanding of the parameters used) with enough complexity, to reflect the possibility of differing communication strategies.

As described in the previous paper,  we define four categories of the party propaganda messages. The first division is between the  ``internal" messages (addressed to the party supporters) and ``external" ones (addressed outside the current support base.
Such a split is quite common in real societies. For example, it is well documented in the case of the conservative/liberal media and their recipients in the United States \cite{adamic05-1,baldassari07-1,stroud10-1,prior12-1,knobloch2012selective}.
We note that the level of polarization of the media in Poland is, probably, even deeper than in the US: there are newspapers, journals, TV stations and WEB pages ``serving" only one group of loyal subscribers. 
Despite much smaller volume of the research devoted to the Polish political behavior, the effects of bias, motivated reasoning and echo chambers were confirmed in an analysis related to the 2010 presidential elections \cite{tworzecki2014knowledge}.
The positive feedback between the perception and confirmation biases of the individual consumers of information and  the economic and political interests of the media companies drives this process relentlessly. In fact, most of the political communication is addressed internally, not externally.
It is only in cases such as the rise of Mr Kukiz campaign, that the media effectively break the barriers and communicate outside the usual envelope, because the rise of his popularity became ``news" in itself.

The media messages may be further divided according to their intent and content. 
The internal propaganda may take two forms: ``mobilizing" (aimed at increasing the emotional commitment of the supporters, transitioning them from the calm into the agitated state, in which they are immune to any contrary argumentation, rational or not); and ``demobilizing", which act in the opposite way: they make the agitated agents calm, and on top of that, they make the calm agents bored, turning them into the neutral state. 
As one can guess, this division is intended to mimic the main propaganda strategies of PiS and PO. 

The external propaganda comes again in two types: ``rational" -- aimed at converting neutral agents into the party supporters, and, additionally, converting calm supporters of other parties into neutral agents via rational argumentation. The latter effect may, in some cases, backfire, as with some probability  the agent receiving the message containing contrary views may become angered by it and turn from its calm state to an agitated one (just like the effects of an encounter between two calm agents supporting different parties in the E/I/O model).   
The second group are ``external-irrational" messages. Mostly, they are a spillover from the internal mobilizing propaganda, and their effect is twofold: they may change neutral agents directly to the agitated supporters, but they turn calm opponents into agitated ones -- as a simple reaction to the emotionally loaded content intended against them. 

A we have noted, the longstanding dominance and the relative stability of the PO/PiS almost symmetric duopoly is interesting because the two parties relied on asymmetric communication strategies. PO has focused on the achievements of the PO government and of Poland as a country: the international position, the economic growth, new infrastructure built with EU funds etc. Most of these achievements were true -- but had little in common with the individual perceptions of the voters, and did not evoke substantial emotional response.
As a result, even minor inconveniences caused by the government policies were sufficient to diminish the commitment of the supporters. Moreover, any misbehavior of the PO politicians opened way to successful attacks.

In contrast, the PiS campaign throughout the period 2005--2007 was almost wholly based on emotional appeal, largely focused on negative feelings. After the lost elections in 2007, and especially after the tragic plane crash which caused the death of the PiS backed president Mr Lech Kaczynski in 2010, almost all communications were directed internally and focused on mobilization of its electorate. 

Due to the separation of the communication channels the situation was remarkably stable, with the exception of 2011, when a splinter party (Twoj Ruch, TR) formed by an ex-PO member of parliament, managed to capture almost 10\% of the votes, at the expense of PO (see Figure~\ref{fig:long}). PO and TR shared most of the ideological baseline, but the TR supporters were mostly people who voted previously for PO but were dissatisfied with the day-to-day policies of the party and the government.

The breakdown of the stability observed in 2011, caused by te asymmetry in the emotional state of the supporters of the two dominant parties and a successful ``invasion" by a newcomer, capturing a foothold in the support base of only one of the parties, has been repeated in 2015, at a much larger scale. 
The support for Mr Kukiz and his political movement has very quickly reached over 20\%, with a corresponding drop in the declared support for PO (Figure~\ref{fig:parties2015}).
The general shape of the support for Mr Kukiz (as known so far) closely resembles that for the TR growth and decay of popularity, which suggests similarity of the driving mechanisms.

\section{Model results}

\subsection{Modifications of the model: from qualitative to semi-quantitative results}

The model presented in reference \cite{sobkowicz2015breakdownV2} provided a qualitative agreement with the observed support evolution. As newer poll results appeared in the press, we became tempted to try to transition from the qualitative into a semi-quantitative regime.
\begin{figure}
	\includegraphics[scale=0.3]{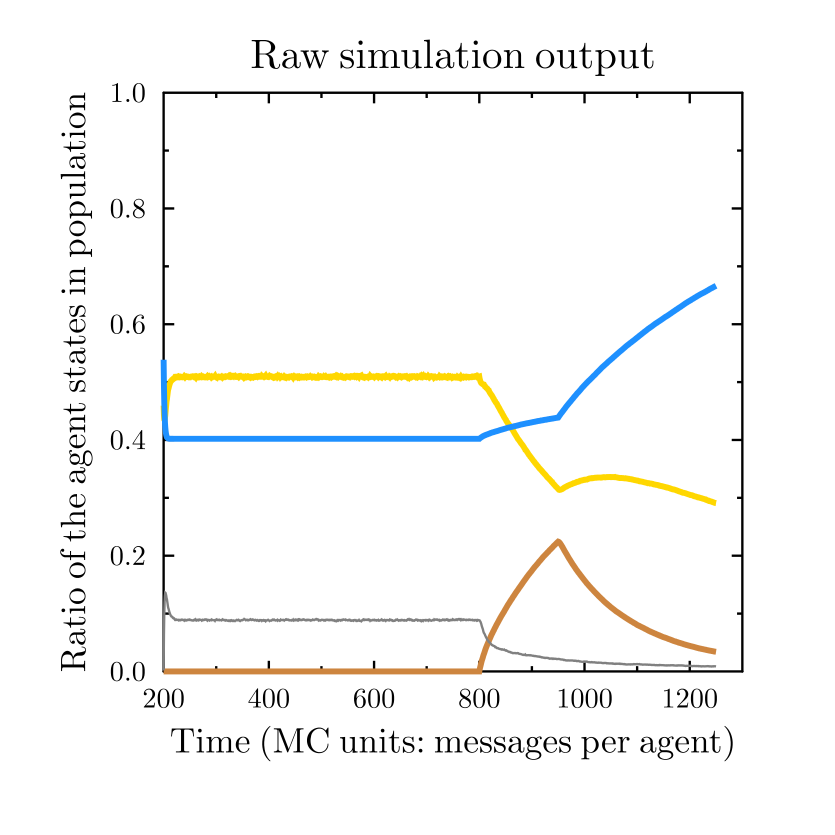}
	\caption{An example of raw (not normalized) results of three party simulation. Orange: PO, blue: PiS, brown: Kukiz, gray: undecided (neutral) agents. Time is measured in MC steps -- messages per agent. The size of the model is 160 000 agents organized in a square, as described in ref~\cite{sobkowicz2015breakdownV2}. \label{fig:raw} }
\end{figure}

The first step in this attempt is to map the results of the three party model onto the real world. Within the model, the support for the three parties and the neutral agents always adds up to 100\%. An example of the raw, unadjusted simulation run results is presented in Figure~\ref{fig:raw}. On the other hand, the reported poll results contain also data for the other parties (which in the Polish case provide a small but non-negligible contribution), as well as the undecided voters (for some of the polls). Using the raw data would naturally lead to a sizable overestimate of the popularity of the three parties.

\begin{figure}
	\includegraphics[scale=1]{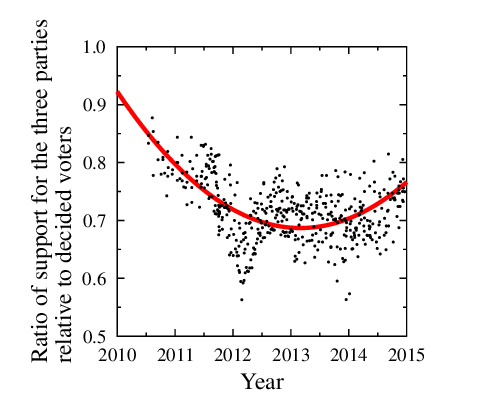}
	\caption{The ratio of the summed support for PiS (together with SP and PJN), PO and Kukiz, in relation to the total reported support for all parties considered within each poll. This allows to normalize the results of the three party agent based model to the real world. Black dots: results for the individual polls, red line: best fit with a quadratic function, used in normalizing the raw simulation results.  \label{fig:normalizing} }
\end{figure}

Fortunately, for the period of 2010--2015, the observed ratio of the summed popularity of the three parties (assuming zero popularity for Mr Kukiz before 2015), has followed a rather stable pattern, which could be approximated by a parabola, as shown in Figure~\ref{fig:normalizing}. 
This allowed us to normalize the simulation results to the reported poll results, by multiplying the raw numbers (summing to 100\%) by the value of the smoothed three-party support in the actual polls. This normalization, which is independent of the simulation parameters allowed to reproduce the observed support quite well, as shown in Figure~\ref{fig:model}.

\subsection{Comparison of simulation results with observations and model predictions}

\begin{table*}
	\begin{tabular}{|p{3.5cm}|p{3cm}|p{3cm}|p{3cm}|}
		\hline
		& Time period A   & Time period B       & Time period C         \\ \hline
		MC steps           & 200--800        & 800--950            & 950--1180             \\ \hline
		Corresponding real time          & before March 14 & March 14 -- June 10 & June 10 -- October 30 \\ \hline \hline
		\textbf{	Message type }      & \multicolumn{3}{c}{\textbf{Message ratio}}   \\ \hline 
		PO mobilizing      &   $0.00$        &   $0.00$            & $0.173\pm 0.014$ \\ \hline
		PO demobilizing    &   $0.10$        &   $0.05$            & $0.017\pm0.024$   \\ \hline
		PO rational        &   $0.25$        &   $0.15$            & $0.108\pm0.030$       \\ \hline
		PO irrational      &   $0.00$        &   $0.00$            & $0.00$             \\ \hline
		\textbf{PO total } &   \bf{0.35}     &   \bf{0.20}         & \bf{0.30}      \\ \hline
		PiS mobilizing     &   $0.40$        &   $0.30$            & $0.294\pm0.024$        \\ \hline
		PiS demobilizing   &   $0.00$        &   $0.00$            & $0.00$                \\ \hline
		PiS rational       &   $0.00$        &   $0.01$            & $0.064\pm0.036$       \\ \hline
		PiS irrational     &   $0.05$        &   $0.09$            & $0.042\pm0.018$        \\ \hline
		\textbf{	PiS total} &   \bf{0.45}     &   \bf{0.40}         & \bf{0.40}       \\ \hline
		Kukiz mobilizing   &   $0.00$        &   $0.10$            & $0.032\pm0.010$          \\ \hline
		Kukiz demobilizing &   $0.00$        &   $0.00$            & $0.018\pm0.010$          \\ \hline
		Kukiz rational     &   $0.00$        &   $0.06$            & $0.012\pm0.010$           \\ \hline
		Kukiz irrational   &   $0.00$        &   $0.04$            & $0.038\pm0.010$         \\ \hline
		\textbf{Kukiz total}   &    \bf{0.00}    &   \bf{0.20}         & \bf{0.10}        \\ \hline
	\end{tabular} 
	\caption{Parameters used in the simulations presented in Figure~\ref{fig:model}.
		for time period C we provide an average value and 2$\sigma$ range of the parameter sets used to study possible variants of parties' communication strategies between July and October 2015.   \label{tab:param} }
\end{table*}

The goals of the current work are twofold. The first is to determine model parameters (especially the parameters describing the party propaganda strategies) which 
lead to a good agreement with the real world data (up to mid-June 2015), and, at the same time, remain in a reasonable agreement with the estimations of the actual propaganda intensities.

The second goal was to study the effects of possible future strategies, 
chosen by the parties  on the expected results of the parliamentary elections scheduled for late October.

The parameters used in the model are designed to correspond to some features of reality, so that, in principle they could be compared with certain characteristics of the social system (such as the relative frequencies of media messages of the four types) or of the individual perception and response. 
Unfortunately, in this quick communication, there is no such comparison, backed by objective data. 
The ranges of the parameters are instead based in the author's estimates of the actual social activity in Poland. 

The first of these is that external sources (media, propaganda) play a more important role in shaping the opinions and emotions related to the political issues in Poland than personal contacts. 
in the workplace and family relationships, in many cases, people avoid the sensitive topic, in order to minimize the conflicts. On the other hand, the media thrive on the polarized messages. For these reasons we have assumed in the simulations that for each agent, on average, 80\% of the political messages come from party propaganda sources, while 20\% are result of interactions with social neighbors.  The 20\% of the messages between the agents are generated directly from the internal states of the agents: their emotions and opinions.

The remaining 80\% of messages, attributed to propaganda, are divided between the three parties, and within these divisions, into the four categories: internal mobilizing, internal demobilizing, external rational and external irrational. At each simulation step, a message is generated randomly, following the probabilities which are the main simulation parameters for the model. These probabilities may, of course, vary in time as the parties' media strategies change.

\begin{figure*}
	\includegraphics[scale=0.8]{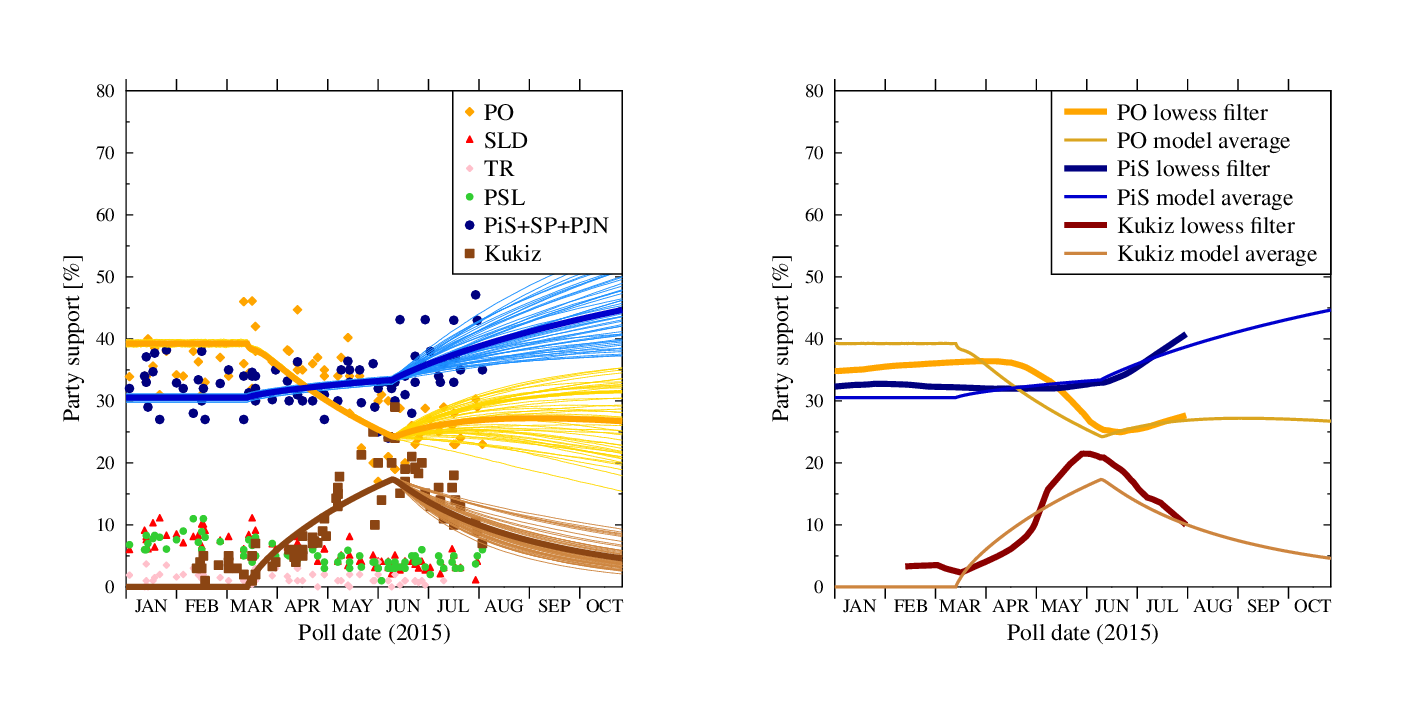}
	\caption{Comparison of the model results with the observed popularity. The model results are extended to October 25th, located at the right edge of the figure panels. Left panel: thin lines: results of simulations for individual communication strategy parameters, thick lines: averaged values.
		Results of the three party agent based model are rescaled to correspond to the actual support in a multi-party environment. Two time points are set as the event timelines: $T1=$ 800 MC steps per agent (rescaled to March 14, 2015) as the appearance of Mr Kukiz propaganda and $T2=$ 950 MC steps (rescaled to late June 10, 2015, second round of the presidential elections), as the time when the three parties change their communication strategies.   The right panel shows, for clarity, a comparison between the smoothed values of the poll results and the average of the 50 runs of the model. \label{fig:model} }
\end{figure*}

To simplify the situation we have split the simulated timespan into four periods. 
The first, up to 200 messages per agent in MC units, is a starting phase, during which the domains of agents holding similar views form (see \cite{sobkowicz2015breakdownV2}). From our point of view this period may be considered as devoted to create the realistic, nonrandom starting conditions.

The next period (denoted as A in Table~\ref{tab:param}), between 200 MC steps and 800 MC steps, is the period during which the effects of the asymmetric propaganda of PO and PiS take effect. As can be seen from Figure~\ref{fig:raw}, for the chosen parameters,  after a very short adjustment period just after T=200, the configuration remains frozen, with the ratio of PO/PiS supporters determined by the distribution of the initial seeds. As shown in Table~\ref{tab:param}, during this period PO strategy is composed mostly (25\% points from the total of 80\% available to both parties) on rational, external propaganda, focused on promoting the country's successes. We assumed that 10\% of PO messages had internal demobilizing effects, as the main storylines did not relate to personal experiences of the party supporters. In contrast, most (40\%) of the messages of PiS were internal mobilizing, as the party focused on its hardcore electorate, stirring and keeping high emotional commitment. The remaining 5\% was assumed to be of  the external irrational  type, due mostly to the spillover of internal propaganda to outsiders, who perceived it as hate and aggression.
We assumed slightly greater value of overall PiS activity, compared with PO, corresponding qualitatively to the actual activity.

The third period (denoted as B in Table~\ref{tab:param}), starting at T=800 MC steps per agent, which coincides with the moment when Mr Kukiz is noticed as one of the presidential candidates and starts to receive media attention. The abstract T=800 MC was thus set to correspond to March 14, 2015. 
The appearance of Mr Kukiz as ``news" naturally decreased the attention given to the two other parties. Moreover, to reflect a relatively lackluster campaign of the PO presidential candidate,
the activity of PO was assumed to drop much more than that of PiS.
PO messages remained split between internal demobilizing and external rational, while PiS largely preserved the focus on internal mobilizing and external irrational. As the campaign entered later stages, at least some of the PiS propaganda became addressed outside the core support base in a rational way, reflected by a small value of the corresponding parameter.
media communications related to Mr Kukiz (assumed to reach 20\% of the media stream) was split between mobilization of the newly won supporters (10\%), some (6\%) rational messages, designed to win new supporters and a smaller fraction (4\%) of messages perceived as irrational. This phase of the simulations is assumed to end at T=950 MC steps, which we set to correspond to June 10, at which time the results of the presidential elections won by the PiS candidate were absorbed by the population and Mr Kukiz begun in earnest to form his political movement.

We have run multiple simulations and combinations of the parameters for the periods A and B, but in this work we present only the results for a single  set which reproduces well the observed changes in the parties popularity till June 2015. 
In contrast, the fourth period (denoted as C), starting with T=950 MC steps, equivalent to June 10, is treated as the basis for prediction of possible outcome of the parliamentary elections
and covers various scenarios of different mixes of each party message types.
 The general choice of the parameters for this period is based on the observed significant changes in the actual media strategies of the parties, visible so far. Let us start with Mr Kukiz: after the initially high media coverage during the presidential campaign and shortly after, the stream of news has dropped sharply. Moreover, Mr Kukiz begun, almost immediately, to quarrel with some of his backers. To reflect this we have decreased his part of the media messages to 10\%, out of which roughly half were internal and half external. A part of the internal messages continued too mobilize, but a significant fraction should be considered demobilizing. Also the fraction of messages externally perceived as irrational has increased sharply, with respect to the total ratio of messages related to Mr Kukiz and his movement.

In contrast, the PiS strategy has changed only slightly -- but with significant results. As before, most of the messages are still addressed to the current electoral base with mobilizing effects. But the composition of the external messages (addressed to the supporters of PO and Mr Kukiz and to the undecided) are now rational and calm, capable of winning new supporters. 
The politicians chosen as campaign leaders are  perceived as conciliatory, calm, rational and positive. The controversial, loaded topics, so prominently visible in the past, are downplayed or absent.  This is reflected, in our simulations, by a much larger probability of external rational messages.

The largest change may be observed in the PO strategy: woken up by the loss of the presidency, the party increased its activity, and while still large effort is directed externally with rational appeals, internally there is a strong shift towards mobilizing messages (based mostly on the negative emotions connected with `what would happen should PiS win').

For the fourth simulation period  we have run 50 simulations with slightly different values of the media parameters. The last column in Table~\ref{tab:param} presents the average values together with the 2$\sigma$ range of the variation between the simulation runs. The evolution of the support for each of such runs is shown as thin line in the left panel of Figure~\ref{fig:model}. 
Despite the relatively low variance of the parameters, the range of the predicted results for October 25 (corresponding to the right edge of the figure) remains substantial. Some simulations (in which PiS uses a particularly effective communication strategy and  PO a very weak one) lead to a 55\%/15\% ratio of votes, giving PiS full majority allowing constitutional changes. At the other extreme, there are simulation in which both parties receive about 30\% of the votes. The thick lines in the left panel of the Figure~\ref{fig:model} correspond to the average over the ensemble of the simulation results, indicating about 45\% votes for PiS, 27\% for PO and about 5\% (which is the threshold for the parliamentary seats) for Mr Kukiz. The right panel in Figure~\ref{fig:model} presents a comparison between the averages of the simulation ensemble and the lowess smoothed poll results, showing a quantitative agreement between the two, up until August 4th.

\subsection{Conclusions}
As we have noted in \cite{sobkowicz2015breakdownV2}, the model on which we base our current results contains significant simplifications and omits quite a few important social phenomena. 
Additionally, the choice of the range of parameters (such as the overall importance of the media or the relative fractions of various types of messages) is based on the personal perception of the author. On the other hand, the E/I/O model has already been used to successfully describe quantitatively the behavior of a different social system, namely participants in an Internet discussion forum \cite{sobkowicz13-3}.  The semi-quantitative agreement of the model results with the observations so far is quite encouraging. 
The comparison of the predictions with future observations might allow us to narrow down the range of the parameters. As we have noted, a careful analysis of the Polish media could provide and independent check on the plausibility  of the model parameters, potentially leading to a cross-check of the model validity or its deficiencies.

\end{document}